\def\gp{\gamma_+}
\def\gm{\gamma_-}

\def\gcm{\gamma_{cm}}
\def\gc{\gamma_c}
\def\bcm{\beta_{cm}}
\def\bc{\beta_c}

\def\ugr{\lower4pt \hbox{$\buildrel > \over \sim$}}
\def\ukl{\lower4pt \hbox{$\buildrel < \over \sim$}}

\def\gpm{\gamma_{\pm}}

\def\bpm{\beta_{\pm}}

\def\ecm{\epsilon_{cm}}
\def\ee{\epsilon_1}
\def\ez{\epsilon_2}

\def\st{\sigma_T}
\def\apm{a_{\pm}}
\def\bpm{b_{\pm}}
\def\cpm{c_{\pm}}
\def\dpm{d_{\pm}}

\documentstyle[epsf, rotate]{laa}
\begin{document}

\thesaurus{}
\title{The Pair Production Spectrum from Photon-Photon annihilation}
\author{M.  B\"ottcher \and R. Schlickeiser}
\institute{Max-Planck-Institut f\"ur Radioastronomie, Postfach 20 24, 53
010 Bonn, Germany}
\date{Received 18 November 1996; Accepted }
\offprints{M. B\"ottcher}

\maketitle

\begin{abstract}

We present the first completely analytical computation of 
the full differential $\gamma$-$\gamma$ pair production
rate from compact radiation fields, exact to 2nd order QED, and
use this result to investigate the validity of previously known 
approximations.

\end{abstract}

\section{Introduction}

The discovery of high-energy $\gamma$-radiation from extragalactic
compact objects has motivated many authors to consider the effects
of $\gamma$-ray absorption by $\gamma$-$\gamma$ pair production,
eventually inducing pair cascades. The relevance of $\gamma$-$\gamma$
pair production to astrophysical systems has first been pointed
out by Nikishov (1962). The first investigation of the $\gamma$-$\gamma$
absorption probability of high-energy photons by different
soft photon fields, along with some useful approximations,
can be found in Gould \& Schr\'eder (1967).

The energy spectrum of injected electrons and positrons due 
to this process has been studied by several authors (e. g.,
Bonometto \& Rees 1971, Aharonian et al. 1983, Zdziarski 
\& Lightman 1985, Coppi \& Blandford 1990). In most 
astrophysically relevant cases, simple approximations 
can be used for this purpose, without much loss of 
accuracy. These usually rely on the high-energy photon 
having much higher energy than the soft photons and
thus dominating the energy input and determining the
direction of motion of the center-of-momentum frame
of the produced pairs. Bonometto \& Rees (1971) used 
basically the same technique as we do, but restricted
their analysis to the case $\ee \gg \ez$, and did not
solve the problem analytically. Two recipes to calculate 
the full energy-dependence of the injected pairs have been 
published (Aharonian et al. 1983 and Coppi \& Blandford
1990), but here the reader is still left with integrations
to be carried out numerically.

It is the purpose of this paper to derive the full energy-spectrum
of pairs, injected by $\gamma$-$\gamma$ pair production, exact to
second order QED for the case of isotropic radiation fields. 
In Section 2, we give a short overview of the kinematics which 
are used in Section 3 to calculate the pair injection spectrum.
In Section 4, we compare our results to well-known approximations 
and specify the limitations of the various approximations.
Our analysis is easily generalized to non-isotropic radiation
fields. The derivation presented here is widely analogous to
the derivation of the pair annihilation spectrum, given by
Svensson (1982).

\section{Kinematics}

We consider an isotropic photon field $n_{ph} (\epsilon)$ where
$\epsilon = h \nu / (m_e c^2)$ is the dimensionless photon energy
in a rest frame which we call the laboratory frame. The Lorentz
invariant scalar product of the four-momenta $\underline{\epsilon_{1,2}}$ 
of two photons having energies $\epsilon_{1,2}$ colliding under 
an angle of cosine $\mu = \cos\theta_{\mu}$ in the laboratory 
frame is then given by

\begin{equation}
\underline{\ee} \cdot \underline{\ez} = \ee \ez (1 - \mu) = 2 \, \ecm^2.
\end{equation}
Here, $\ecm$ is the photon energy in the center-of-momentum 
frame. In order to allow for the possibility to create an 
electron-positron pair, conservation of energy implies 
$\ecm = \gcm$, and the condition $\gcm \ge 1$ determines 
the pair-production threshold. $\gcm$ is the 
Lorentz factor of the electron/positron in the cm frame 
where the produced electrons move with speed $\pm \bcm c$
and $\bcm = \sqrt{1 - 1/\gcm^2}$. The definition of the angle
variables needed in this calculation is illustrated in Fig. 1.

\begin{figure}
\epsfysize=4.5cm
\epsffile[70 290 320 530] {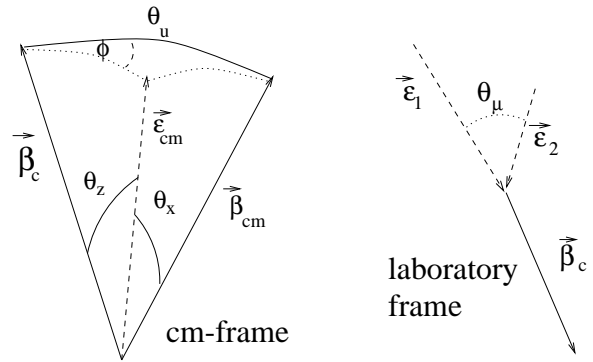}
\caption[]{Definition of the angles in cm and laboratory frame. 
$\epsilon_{1,2}$ denotes the direction of motion of an incoming 
photon, $\beta_{cm,\pm}$ is the direction of motion of the produced 
electron and positron in the cm and the laboratory frame,
and $\bc$ characterizes relative motion of the laboratory and 
the cm-frame, respectively.}
\end{figure}

The cm frame moves relative to the laboratory frame with velocity
$- \bc c$ and Lorentz factor $\gc = (1 - \bc^2)^{-1/2}$. The four
velocity of the laboratory frame ($\bc c$ in the cm frame) is denoted 
by $\underline{v_L}$. The Lorentz factors of the produced pairs in the 
laboratory frame are related to the cm quantities by

\begin{equation}
\gpm = \gcm \gc \, (1 \pm \bcm \bc u). 
\end{equation}
Evaluating the Lorentz invariant scalar product

\begin{equation}
\underline{\epsilon_{1,2}} \cdot \underline{v_L} = \epsilon_{1,2} 
= \ecm \gc (1 \pm \bc z)
\end{equation}
in the laboratory and the cm-frame, respectively, we find

\begin{equation}
\gc = {E \over 2 \, \ecm}, \hskip 1cm {\rm where} \hskip 0.8cm E = \ee + \ez,
\end{equation}
and

\begin{equation}
z = cos\theta_z = {\ee - \ez \over N}, \hskip 0.6cm {\rm where} \hskip 0.4cm
N = \sqrt{E^2 - 4 \, \ecm^2}.
\end{equation}
Inserting Eq. (4) into Eq. (2) and using energy conservation ($\ecm = \gcm$)
fixes the angle cosine $u$ to

\begin{equation}
u = u_0 \equiv {E - 2 \, \gm \over \bcm \, N}.
\end{equation}
The differential cross section for $\gamma$-$\gamma$ pair production
(see Eq. [11]) depends on

\begin{equation}
x = cos\theta_x = u z + \sqrt{1 - u^2} \sqrt{1 - z^2} \, \cos\phi.
\end{equation} 

\section{The pair yield}

The differential yield of produced pairs is calculated as

$$
\dot n (\gm) $$
\begin{equation}
= {c \over 4} \int\limits_0^{\infty} d\ee \, n_{ph} (\ee) 
\!\!\!\!\!\!\!\!\!\!\!\!\!\!\!\!\! \int\limits_{\max\left\lbrace 
{1\over\ee}, \, \gm + 1 - \ee \right\rbrace}^{\infty} 
\!\!\!\!\!\!\!\!\!\!\!\!\!\!\!\!\! d\ez \, n_{ph} (\ez) \!\!\!
\int\limits_{-1}^{1 - {2 \over \ee \ez}} d\mu \> (1 - \mu) 
\, {d\sigma \over d\gm}
\end{equation}
where
\begin{equation}
{d\sigma \over d\gm} = \oint d\Omega_{cm} {d^2 \sigma \over
d\Omega_{cm} d\gcm} {d\gcm \over d\gm}.
\end{equation}
The differential cross section has been evaluated by
Jauch \& Rohrlich (1959):

\begin{equation}
{d^2\sigma \over d\Omega_{cm} d\gcm} = \delta(\ecm - \gcm) {d\sigma
\over d\Omega_{cm}},
\end{equation}
where
$$
{d\sigma \over d\Omega_{cm}} = {1 \over 2\pi} {3 \over 16} \st
{\bcm \over \ecm^2} \cdot 
$$
$$
\cdot \, \Biggl\lbrace -1 + {3 - \bcm^4 \over 2} \left( {1 \over 
1 - \bcm x} + {1 \over 1 + \bcm x} \right) 
$$
\begin{equation}
\hskip 1.3cm
- {1 \over 2 \, \ecm^4} \left( {1 \over [1 - \bcm x]^2} + {1 \over 
[1 + \bcm x]^2} \right) \Biggr\rbrace.
\end{equation}
We may express the solid angle element $d\Omega_{cm} = d u \, d\phi$. 
Using Eq. (2), we find

\begin{equation}
\delta (\ecm - \gcm) \, {d\gcm \over d\gm} = 
\delta(u - u_0) \, {2 \over N \, \bcm}.
\end{equation}
This enables us to carry out the $u$-integration in
Eq. (9) immediately. If we write the denominators in 
Eq. (11) as
\begin{equation}
1 \pm \bcm x = \apm + \bpm \cos\phi
\end{equation}
with
\begin{equation}
\apm \equiv 1 \pm u_0 \, z \, \bcm, \hskip 0.8cm
\bpm \equiv \pm \sqrt{1 - u_0^2} \sqrt{1 - z^2} \bcm,
\end{equation}
we find

$$
{d\sigma \over d\gm} = {3 \over 8} \st {1 \over N \, \ecm^2}
\cdot 
$$
$$ \cdot {1 \over 2\pi} \int\limits_0^{2\pi} d\phi \Biggl\lbrace -1 
+ {3 - \bcm^4 \over 2} \left( {1 \over a_+ + b_+ \cos\phi}
+ {1 \over a_- + b_- \cos\phi} \right) 
$$
$$ - {1 \over 2 \, \ecm^4}
\left( {1 \over [a_+ + b_+ \cos\phi]^2} + {1 \over [a_- + b_- \cos\phi]^2}
\right) \Biggr\rbrace 
$$
$$
= {3 \over 8} \st {1 \over \ecm} \cdot 
$$
\begin{equation}
\cdot \left\lbrace - {1 \over N \, \ecm} + {3 - \bcm^4 \over 4} 
\left( G_+ + G_- \right) - {1 \over 8 \, \ecm^2} \left( F_+ + F_- \right) 
\right\rbrace
\end{equation}
where

\begin{equation}
G_{\pm} = {1 \over \sqrt{\ee \ez + \ecm^2 \cpm}},
\end{equation}

\begin{equation}
F_{\pm} = {\dpm - 2 \, \ecm^2 \over \left( \ee \ez + \ecm^2 \cpm 
\right)^{3/2}},
\end{equation}
with

\begin{equation}
\cpm \equiv \left( \epsilon_{1,2} - \gm \right)^2 - 1,
\end{equation}

\begin{equation}
\dpm \equiv \epsilon_{1,2}^2 + \ee \ez \pm \gm \, (\ez - \ee)
\end{equation}
and we used the integrals

\begin{equation}
\int\limits_0^{2\pi} {d\phi \over a + b \cos\phi} =
{2\pi \over \sqrt{a^2 - b^2}},
\end{equation}

\begin{equation}
\int\limits_0^{2\pi} {d\phi \over (a + b\cos\phi)^2} =
{2 \pi \, a \over (a^2 - b^2)^{3/2}}
\end{equation}
and the identity

\begin{equation}
\apm^2 - \bpm^2 = 4 {\ee \ez +\ecm^2 \, \cpm \over N^2 \ecm^2}
\end{equation}
which follows from Eqs. (5), (6) and (14). Now, 
inserting Eq. (15) into Eq. (8)  yields the exact expression
for the differential pair injection rate. Using Eq. (1) we transform
the $\mu$ integration into an integration over $d\ecm$. This leads
us to

$$
\dot n (\gm) = {3 \over 4} \st \, c \int\limits_0^{\infty} d\ee
\> n_{ph} (\ee)  \cdot
$$
$$ \cdot \int\limits_{\max\left\lbrace {1 \over \ee}, \, 
\gm + 1 - \ee \right\rbrace}^{\infty} d\ez \> n_{ph} (\ez) \, 
{1 \over (\ee \ez)^2}\int\limits_{\ecm^L}^{\ecm^U} d\ecm \> \ecm^2 
\cdot 
$$
\begin{equation}
\cdot \left\lbrace - {1 \over N \, \ecm} + {3 - \bcm^4 \over 4} 
\left( G_+ + G_- \right) - {1 \over 8 \, \ecm^2} \left( F_+ + F_- \right) 
\right\rbrace
\end{equation}
which can be calculated analytically. The integration limits follow from
$1 \le \ecm \le \sqrt{\ee\ez}$ and the condition $\gm \lower4pt 
\hbox{$\buildrel < \over >$} \gcm \gc (1 \pm \bcm \bc)$ which yields

\begin{equation}
\ecm^U = \min\left\lbrace \sqrt{\ee\ez}, \, \ecm^{\ast} \right\rbrace,
\hskip 0.8cm
\ecm^L = \max\left\lbrace 1, \, \ecm^{\dag} \right\rbrace
\end{equation}
where

$$
\left(\ecm^{\ast, \dag}\right)^2 = 
$$
\begin{equation}
{1 \over 2} \left( \gm [ E - \gm] + 1 \pm 
\sqrt{ (\gm [ E - \gm] + 1)^2 - E^2} \right).
\end{equation}
Using the integrals 2.271.4, 2.271.5, 2.272.3, 2.272.4, 
and 2.275.9, of Gradshteyn \& Ryzhik (1980), we find
as final result for the differential pair yield

$$ \dot n(\gm) = {3 \over 4} \st \, c \int\limits_0^{\infty} d\ee 
\> { n_{ph} (\ee) \over \ee^2} \!\!\! \int\limits_{\max\left\lbrace 
{1 \over \ee}, \, \gm + 1 - \ee \right\rbrace}^{\infty} \!\!\!\!\!\!\!\!
d\ez \> {n_{ph} (\ez) \over \ez^2} \, \cdot $$
\begin{equation} 
\hskip 1.5cm \cdot \> \left\lbrace {\sqrt{ E^2 - 4 \, \ecm^2} 
\over 4}  + H_+ + H_- \right\rbrace \Biggr\vert_{\ecm^L}^{\ecm^U}, 
\end{equation}
where for $\cpm \ne 0$ we have

$$ H_{\pm} = - {\ecm \over 8 \, \sqrt{\ee \ez + \cpm \ecm^2}} \left( 
{\dpm \over \ee \ez} + {2 \over \cpm} \right) 
$$
$$
+ {1 \over 4} \left( 2 - {\ee \ez
- 1 \over \cpm} \right) \, I_{\pm}
$$
\begin{equation}
\hskip 1.5cm
+ {\sqrt{\ee\ez + \cpm \ecm^2} \over 4} \left( {\ecm \over
\cpm} + {1 \over \ecm \ee \ez} \right) \end{equation}
and

$$
I_{\pm} = 
$$
\begin{equation}
\cases{ {1 \over \sqrt{\cpm}} \ln \left( \ecm \sqrt{\cpm} + 
\sqrt{\ee \ez + \cpm \ecm^2} \right) & for $\cpm > 0$ \cr
{1 \over \sqrt{- \cpm}} \arcsin \left( \ecm \sqrt{ - {\cpm \over
\ee \ez}} \right) & for $\cpm < 0$. \cr}
\end{equation}
For $\cpm = 0$ we find

$$ 
H_{\pm} = \left( {\ecm^3 \over 12} - {\ecm \dpm
\over 8} \right) {1 \over (\ee \ez)^{3/2}}
$$
\begin{equation}
\hskip 2cm 
+ \left( {\ecm^3 \over 6} + {\ecm \over 2} + {1 \over 4 \ecm} 
\right) {1 \over \sqrt{\ee \ez}} .
\end{equation}

\section{Comparison to approximations}

Now, we use the exact expression, given in Eq. (26) to
specify the regimes of validity and the limitations of 
various approximations. The first detailed computation
of the pair production spectrum was presented by Bonometto
\& Rees (1971). Based on the neglect of the energy
input of the soft photon, they basically follow the
same procedure as described above, but do not carry
out the angle-integration (integration over $\ecm$ in
our formalism) analytically. In the case $\ee \gg \ez$,
it is in very good agreement with the exact result, but
its evaluation is even more time-consuming than using
the latter. For this reason, we will not consider it
in detail, but concentrate on approximations which
really yield simpler expressions than the exact one.

\subsection{$\delta$-function approximation for power-law 
spectra}

Probably the simplest expression for the pair
spectrum injected by $\gamma$-rays interacting 
with a power-law of energy spectral index 
$\alpha' = \alpha - 1$ ($n(\ez) \propto \ez^{-\alpha}$) 
is based on the assumption $\ee \gg \ez$ and on the 
well-known fact that photons of energy $\ee$ interact 
most efficiently with photons of energy $\ez = 2/\ee$ 
which motivates the approximation for the 
$\gamma$-$\gamma$ opacity of Gould \& Schr\'eder 
(1967) using a $\delta$ function approximation
for the cross section, 
$$ \sigma_{\gamma\gamma} (\ee, \ez) \approx 
{1\over 3} \st \ez \delta\left( \ez - 
{2 \over \ee} \right) $$ 
Since in this approach pair production 
takes place only near the pair
production threshold ($\gcm \approx 1$), the
produced pairs have energies $\gp \approx \gm
\approx (\ee + \ez)/2$ (Bonometto \& Rees 1971).
The resulting pair injection spectrum is therefore

\begin{equation}
\dot n (\gamma) \approx \eta(\alpha') \, c \, \st \, 
{ n_{ph} (\epsilon_0) \over \epsilon_0} \, n_{ph} 
\left( {1 \over \epsilon_0} \right)
\end{equation}
(e. g. Lightman \& Zdziarski 1987)
where $\epsilon_0 \equiv \gamma + \sqrt{\gamma^2 - 1}$
(which, of course, reduces to $2 \gamma$ for $\gamma \gg 1$)
and $\eta(\alpha')$ is a numerical factor, depending
only on the spectral index of the soft photon distribution.
This approximation yields useful results, if the power-law
photon spectra extend over a sufficiently wide range
($\epsilon_U / \epsilon_L \ugr10^4$) and if for every
high-energy photon of energy $\ee$ there is a soft photon
of energy $\ez = 2/\ee$. Else, the injection spectrum
calculated with Eq. (30) cuts off at the inverse of
the respective cutoff of the soft photon spectrum, seriously
underpredicting the injection of pairs of higher or lower energy,
respectively, where the injection spectrum declines smoothly.
Nevertheless, these pairs can still carry a significant fraction 
of the injected power. Eq. (30) fails also to describe 
the injection of low-energetic pairs in case of a high
lower cut-off of the $\gamma$-ray spectrum even if
soft photons of energy $\ez = 1/\ee$ are present. For
example, in the case of the interaction of a power-law 
spectrum extending from $\ee = 10^2$ -- $10^6$ with a 
soft power-law spectrum extending from $\ez = 10^{-7}$ 
-- $10^{-2}$ Eq. (30) overpredicts the injection 
of pairs slightly above $\gamma = 50$ by an order of 
magnitude and cuts off below this energy. A similar
problem arises at the high-energy end of the injection
spectrum. An example for this fact is shown in
Fig. 2. In contrast, the approximation (30) can
well be used to describe the injection of pairs at
all energies if both photon fields extend up to 
(and down to, respectively) $\epsilon_{1,2} \sim 1$.
For more general soft photon distributions which are 
different from a power-law (e. g. a thermal spectrum) 
the analogous $\delta$-function approximation has first 
been introduced by Kazanas (1984). 

\begin{figure}
\rotate[r] {
\epsfxsize=6cm
\epsffile[100 20 600 50] {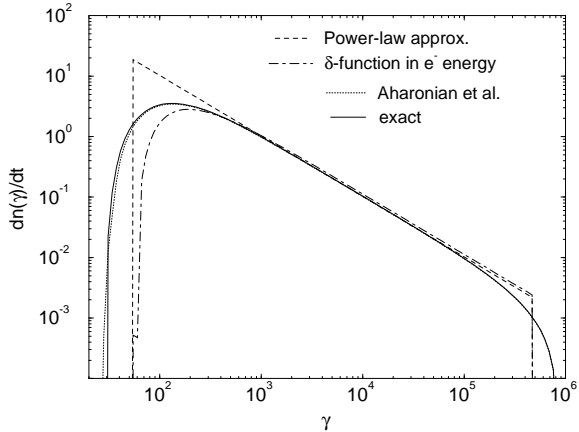} }
\caption[]{Differential pair injection rate (arbitrary 
units) for the interaction of a power-law from $\ee =
10^2$ -- $10^6$, photon spectral index $\alpha_{\gamma} 
= 1.5$, with a soft power-law from $\ez = 10^{-7}$ -- $10^{-2}$,
$\alpha_s = 1.5$. }
\end{figure}

\subsection{$\delta$-function in electron energy}

Using the full cross section for $\gamma$-$\gamma$
pair production as given by Jauch \& Rohrlich (1959)
instead of the $\delta$-function approximation
adopted in Eq. (30) does not reduce the limitations
of this power-law approach significantly, but other
soft photon distributions can be treated more
successfully with this approximation which in the
limit $\ee \gg \ez$ reads

\begin{equation}
\dot n(\gamma) \approx 2 \, c \, { n_{ph}^{(1)} 
(2\gamma) \over (2\gamma)^2} \!
\int\limits_{1/2\gamma}^1 \!\! d\ez \> 
{n_{ph}^{(2)} (\ez) \over (\ez)^2} \!\!
\int\limits_{(\ecm^L)^2}^{(\ecm^U)^2} \!\! 
ds \> s \, \sigma_{\gamma\gamma} (s)
\end{equation}
where $s = \ecm^2$ and the limits $\ecm^{U/L}$ 
are given by Eqs. (24) and (25). Here, we have
assumed that the produced electron and positron 
have energy $\gm = \gp = \ee/2$. This 
approach works equally well for power-law photon 
fields, but in contrast to Eq. (30), it tends to
underpredict the injection of low-energetic
pairs. The same is true for the interaction
of $\gamma$-ray photon fields with thermal
soft photon fields where the high-energy tail
of the injection spectrum is described very
accurately (a few~\% error) by Eq. (31).
The accuracy of this approximation improves
with decreasing lower cut-off of the $\gamma$-ray
spectrum. E. g., the injection due to a power-law
$\gamma$-ray spectrum from $\ee = 2$ -- $10^6$
interacting with a thermal spectrum of normalized
termpature $\Theta = 0.1$ is described by Eq. (31)
with a deviation of less than 30~\% from the exact
result down to $\gamma \sim 1$. For $\gamma > 2$,
the deviation was much less than 10~\%. We show
an example for the latter situation in Fig. 3.

\begin{figure}
\rotate[r] {
\epsfxsize=6cm
\epsffile[100 20 600 50] {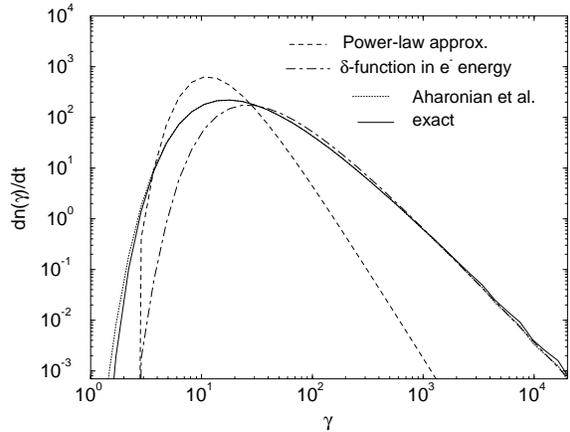} }
\caption[]{Differential pair injection rate (arbitrary 
units) for the interaction of a power-law $\gamma$-ray
spectrum from $\ee = 5$ to $10^6$, $\alpha_{\gamma} = 1.5$,
with a thermal blackbody spectrum of temperature 
$\Theta = 10^{-2}$. }
\end{figure}

\subsection{Approximation by Aharonian et al.}

A very useful approximation to the pair injection 
spectrum for all shapes of the soft photon spectrum 
under the condition $\ez \ll 1 \ukl \ee$ has been found by 
Aharonian et al. (1983). They use a different representation
of the pair production cross section and end up with a
one-dimensional integral over $k = \sqrt{\ee^2 + \ez^2
+ 2 \ee \ez \mu}$ which is equivalent to our $\ecm$
integration in Eq. (21). They solve this integration
analytically after simplifying the integrand and the 
integration limits according to the assumptions
mentioned above. The resulting injection spectrum is

$$ 
\dot n(\gamma) \approx {3 \over 32} \, c \, \st 
\int\limits_{\gamma}^{\infty} d\ee \> {n^{(1)}_{ph} 
(\ee) \over \ee^3} \int\limits_{\ee \over 4 \gamma 
(\ee - \gamma)}^{\infty} d\ez \> {n^{(2)}_{ph} (\ez) 
\over \ez^2} \cdot 
$$
$$ \cdot \, \Biggl\lbrace {4 \ee^2 \over \gamma (\ee - 
\gamma)} \ln\left( {4 \ez \gamma (\ee - \gamma) \over 
\ee} \right) - 8 \ee \ez 
$$
\begin{equation}
\hskip 1cm
+ {2 (2 \ee \ez - 1) \ee^2 \over \gamma (\ee - \gamma)}
- \left( 1 - {1 \over \ee \ez} \right) {\ee^4 \over
\gamma^2 (\ee - \gamma)^2} \Biggr\rbrace.
\end{equation}
It describes the power-law tail of the pair spectrum
injected by power-law $\gamma$-ray photon fields 
perfectly and is much more accurate to the injection 
of low-energy pairs. Interaction with a power-law
soft photon field is reproduced within errors
of only a few~\%. Even if as well the $\gamma$-ray
as the soft photon spectrum extend to $\epsilon_{1,2}
= 1$, the error at $\gamma \sim 1$ increases only
to $\sim$ 10~\%. Problems with this approximation
arise if the soft photon spectrum extends up
to $\ez \sim 1$, but the $\gamma$-ray spectrum
has a lower cut-off $\ee \gg 1$. In this case,
the injection of low-energetic pairs is seriously
overpredicted by Eq. (32). For power-law soft
photon fields, the integration over $\ez$ in Eq. (32)
can be carried out analytically, as was found by
Svensson (1987). His Equation (B8) (multiplying
with the total absorption coefficient $\propto 
\epsilon^{\alpha}$ by which the total injection 
rate had been normalized to 1) is in perfect 
agreement with the numerical results according to 
Eq. (32).

The interaction of power-law $\gamma$-ray spectra
with thermal soft photon fields is generally described
within an error of a few \% at all electron/positron
energies if the soft photon temperature is 
$\Theta \ukl 0.1$, even if the $\gamma$-ray 
spectrum extends down to $\ee \sim 1$.

\begin{figure}
\rotate[r] {
\epsfxsize=6cm
\epsffile[100 20 600 50] {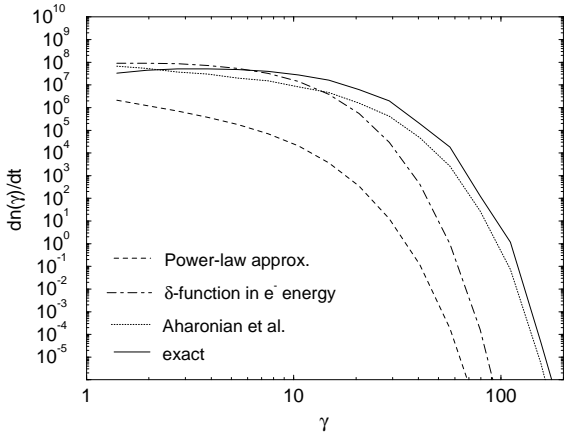} }
\caption[]{Differential pair injection rate (arbitrary 
units) for the interaction of a thermal blackbody spectrum
of temperature $\Theta = 5$ with itself. }
\end{figure}

Interestingly, even the interaction of a mildly 
relativistic thermal photon field ($\Theta \ukl 3$) 
with itself (for which Aharonian's approximation 
was not designed) is reproduced reasonably well, 
but the result of Eq. (32) differs from the exact 
injection rate by a roughly constant factor. When 
artificially introducing a factor adjusting the 
high-energy tails of the injection spectra, Eq. (32) 
overpredicts the injection of low-energetic 
pairs by a factor of $\sim 3$, but for $\gamma \ugr 3$ 
there is very good agreement with the exact result. The
deviation becomes more important with increasing photon 
temperature, and for $\Theta = 5$ the injection of 
cold pairs is already overpredicted by a factor of
$> 10$. Fig. 4 illustrates the accuracy of the various
approximations for a compact thermal radiation of
temperature $\Theta = 5$.

We find that all the statements on soft photon or
$\gamma$-ray power-law spectra made above are only 
very weakly dependent on the respective spectral
index.

\section{Summary and conclusions}

We derived the full energy spectrum of injected pairs,
produced by $\gamma$-$\gamma$ pair production and
compared the result to the various approximations
known before. We found that the simplest expressions,
based on $\delta$ function approximation to the
cross section can well reproduce the power-law
tail of the injection spectrum, but have problems
at low particle energies.

The approximation of Aharonian et al. (1983) was
shown to be the most accurate one and even yields
useful results in regimes for which this approach
was not designed.

\acknowledgements{
We thank the referee, R. Svensson, for very helpful comments.}

\end{document}